\documentclass[12pt]{article} 

\usepackage[utf8]{inputenc} 

\usepackage{amsmath}
\usepackage{graphicx,psfrag,epsf}
\usepackage{enumerate}
\usepackage{natbib}
\usepackage{url} 

\newcommand{\blind}{1}

\RequirePackage{amsthm,amsmath,amsfonts}
\RequirePackage[colorlinks,citecolor=blue,urlcolor=blue]{hyperref}
\usepackage{multirow}
\usepackage{algorithm2e}
\usepackage{comment}

\usepackage{pgfplots}
\usepackage{pgfplotstable}
\usepackage{here}
\usepackage{float}
\usepackage[font=small]{caption}

\usepgfplotslibrary{statistics}

\numberwithin{equation}{section}

\newtheorem{prop}{Proposition}

\newtheorem{proc}{Procedure}
\newtheorem{rem}{Remark}[section]

\newcommand{\norme}[1]{\left\Vert #1\right\Vert}

\newcommand\OU{Ornstein-Uhlenbeck\xspace}
\newcommand{\BRD}{Brownian motion with drift\xspace}

\newcommand{\Pv}{\textit{p}-value\xspace}

\usepackage{titlesec} 

\titleformat{\subsection}[runin]
  {\normalfont\bfseries}{\thesubsection}{1em}{}
  
\usepackage{color}
\usepackage{tikz}
\usetikzlibrary{fadings,shapes.arrows,shadows} 
\tikzfading[name=arrowfading, top color=transparent!0, bottom color=transparent!95]
\tikzset{arrowfill/.style={top color=red, bottom color=red}}
\tikzset{arrowstyle/.style={draw=red,arrowfill, single arrow,minimum height=#1, single arrow,
single arrow head extend=.4cm,}}

\newcommand{\tikzfancyarrow}[2][2cm]{\tikz[baseline=-0.5ex]\node [arrowstyle=#1] {#2};}

\usepackage{fancybox}

\usepackage{xcolor}
\colorlet{lightblue}{blue!10}
\usepackage{amssymb}

\begin{document}

\def\spacingset#1{\renewcommand{\baselinestretch}%
{#1}\small\normalsize} \spacingset{1}


%
%

\if1\blind
{
  \title{\bf A Sequential Algorithm to Detect Diffusion Switching along Intracellular Particle Trajectories}
  \author{V. Briane$^{\star \ddag}$, C. Kervrann$^\star$, M. Vimond$^\ddag$\\
    INRIA, Centre de Rennes Bretagne Atlantique, Serpico Project-Team$^\star$\\
    CREST (Ensai,  Universit\'e Bretagne Loire)}
    \date{}
  \maketitle
} \fi

\if0\blind
{
  \bigskip
  \bigskip
  \bigskip
  \begin{center}
    {\LARGE\bf A Sequential Algorithm to Detect Diffusion Switching along Intracellular Particle Trajectories}
\end{center}
  \medskip
} \fi

\bigskip
\begin{abstract}
\noindent Single-particle tracking allows to infer the motion of single molecules in living cells. When we observe a long trajectory (more than 100 points), it is possible that the particle switches mode of motion over time. Then, fitting a single model to the trajectory can be misleading. In this paper, we propose a method to detect the temporal change points: the times at which a change of dynamics occurs. More specifically, we consider that the particle switches between three main modes of motion: Brownian motion, subdiffusion and superdiffusion. We use an algorithm based on a statistic \citep{briane2016adaptive} 
 computed on local windows along the trajectory. The method is non parametric as the statistic is not related to any particular model. This algorithm controls the number of false change point detections in the case where the trajectory is fully Brownian. A Monte Carlo study is proposed to demonstrate the performances of the method and also to compare the procedure to two competitive algorithms. 
At the end, we illustrate the utility of the method on real data depicting the motion of mRNA complexes --called mRNP-- in neuronal dendrites. 
\end{abstract}

\noindent%
\vfill

\newpage
\spacingset{1.45} 
\section{Introduction}
A cell is composed of lots of structures in interaction with each other. 
They continuously exchange biological material, such as proteins, directly via the cytosol or via networks of polymerised filaments namely the microtubules,  actin filaments and intermediate filaments. The dynamics of these proteins determine the organization and function of the cell \cite[chapter 8]{bressloff2014stochastic}. Then, it is of paramount interest to quantify the main modes of mobility of molecules in living cells.

 In this paper, we consider that the dynamics of molecules can be classified into three groups: subdiffusion, superdiffusion and Brownian motion. 
Usually, in the biophysics literature, the definition of these dynamics is related to the criterion of the mean square displacement (MSD), see for example \citep{qian1991single}. 
Given a particle trajectory $(X_t)_{t>0},$ the MSD is defined as the function of time,
\begin{equation}
\mathrm{MSD}(t)=\mathbb{E}\left(\left\|X_{t+t_0}-X_{t_0}\right\|^2\right),
\end{equation}
where $\|\cdot\|$ is the euclidean norm and $\mathbb{E}(\cdot)$ is the expectation of the probability space. 
If the MSD function is linear ($\mathrm{MSD}(t)\propto t$), the trajectory is Brownian \citep{qian1991single}. 
 If the MSD is sublinear (respectively superlinear) the trajectory is a subdiffusion (respectively a superdiffusion), see \citep[Chapter~7]{bressloff2014stochastic} and \citep{metzler2000random}.
 The biological interpretation of subdiffusion is that the particle is confined in a domain or evolves in an open but crowded area \cite[Section~3]{berry2014anomalous,bressloff2013stochastic}. Superdiffusion occurs when the particle is transported actively via molecular motors along the microtubules \cite[Section~4]{bressloff2013stochastic}. Finally, when the particle evolves freely inside the cytosol, it undergoes Brownian motion \cite[Section~2]{bressloff2013stochastic}.  
\cite{briane2016adaptive} propose a consistent three decision test to distinguish the three types of diffusion.\newline
 In what follows, we use the test statistic proposed in \cite{briane2016adaptive} in a new setting, namely change point analysis. Actually, as intracellular transport presents a high heterogeneity of motions depending on the spatial location, the particle switches dynamics over time while crossing different areas of the cell. 
Let us take the example of the postsynaptic AMPA-type glutamate receptors (AMPARs), a protein involved in the fast excitatory synaptic transmission. These proteins can evolve freely in the synapses and then be trapped in a potential well created by the interaction with an ensemble of other molecules. Then, if we observe the trajectory of an AMPAR protein during a long period, we should see multiple switches between Brownian motion and subdiffusion.  As another example, \cite{lagache2009quantitative} model the dynamics of a virus invading a cell to infer its mean arrival time to the cell nucleus where it replicates. 
In the model of \cite{lagache2009quantitative}, the virus motion alternates between superdiffusion along the microtubules and Brownian motion in the cytosol. To address several issues in motion regime changes, we develop here a sequential method based on a statistic proposed in \cite{briane2016adaptive}  for detecting the time at which an intracellular particle changes dynamic. More precisely, we detect the time at which the particle changes from one type of diffusion (superdiffusion, subdiffusion or Brownian motion) to another type of diffusion.\newline
The present paper is organized as follows.
In Section \ref{sec:model_cp}, we exhibit the inference model. In section \ref{sec:test}, we describe our problem as a statistical test. In Section \ref{sec:proc}, we present our sequential procedure for detecting change points along a trajectory. Section \ref{sec:cut_off} is devoted to the automatic selection of thresholds parameters used in the procedure. We give some simulations results in Section \ref{sec:algo_sim} and compare to other methods in Section \ref{sec:comparison}. Finally, in Section \ref{sec:real_data}, we analyse real data depicting the motion of mRNA complexes --called mRNP-- in neuronal dendrites.

\section{Change Point Model}
\label{sec:model_cp}
We observe the successive positions of a single particle in a $d$-dimensional space ($d=2$ or $d=3$) at time $t_0,t_1,\dots,t_n$. We suppose that the lag time between two consecutive observations is a constant $\Delta$. The observed trajectory of the particle is,
\begin{equation}
\mathbb{X}_n=\left({X_{t_0},X_{t_1},\dots,X_{t_{n}}}\right),
\end{equation}
where $X_{t_k}\in \mathbb{R}^d$ is the position of the particle at time $t_k=t_0+k\Delta$, $k=0,\ldots,n$.
We assume that the discrete trajectory is generated by a $d$-dimensional ($d=2$ or $d=3$) diffusion process $(X_t)$ strong solution of the stochastic differential equation:
\begin{equation}
dX_t =\mu(X_t,t)dt+\sigma(t) dB^{\mathfrak{h}(t)}_t,\qquad t\in[t_0,t_n],
\label{eq:sde2}
\end{equation}
where $B^{\mathfrak{h}(t)}$ denotes a $d-$dimensional fractional Brownian motion of Hurst parameter $\mathfrak{h}(t)$; the unknown parameters of the model are the Hurst parameter function $\mathfrak{h}:\mathbb{R}^+\to(0,1)$, the diffusion coefficient function $\sigma:\mathbb{R}^+\to (0,\infty)$ and the drift term $\mu:\mathbb{R}^+\times \mathbb{R}^d\to\mathbb{R}^d.$  \newline
\indent Furthermore, we assume that there exists a sequence of $N$ change points on $
[t_0,t_n],$ namely $t_0=\tau_0<\tau_1<\ldots\tau_N<\tau_{N+1}=t_n$ such that,
\begin{align}
\forall j\in\{0\ldots N\}, \forall x \in \mathbb{R}^d, \forall t \in [\tau_j,\tau_{j+1}),\quad  
\mu(x,t)&=\mu_j(x)\\
\mathfrak{h}(t)&=\mathfrak{h}_j\\
\sigma(t)&=\sigma_j. 
\end{align}
The number of change points $N$, the drift functions $(\mu_j)_{j=0\ldots N}$ and 
the diffusion coefficient $(\sigma_j)_{j-0\ldots N}$ are unknown. We note that as $N$ is unknown the vector of change points $(\tau_j)_{j=1\ldots N}$ is also unknown. We also assume that the drift terms $\mu_j$ satisfy usual Lipschitz and linear growth conditions, see \citep{nualart2002regularization} for $0<\mathfrak{h}\leq 1/2$ and \citep{mishura2008stochastic} for $1/2<\mathfrak{h}<1$. Then, the stochastic differential equation \eqref{eq:sde2} admits a strong solution on each interval $[\tau_j,\tau_{j+1})$. We extend by continuity the solution on each subinterval to get a solution on $[t_0, t_n]$. Moreover, we assume that $(\mathfrak{h}_j,\mu_j)$ and $(\mathfrak{h}_{j+1},\mu_{j+1})$ are associated to different types of diffusion. We note that the parameter $\sigma_j$ does not influence the type of diffusion. For example, $\mathfrak{h}_j=1/2$ and $\mu_j(x)=0$ define the Brownian motion on $[\tau_j,\tau_{j+1})$ then $(\mathfrak{h}_{j+1},\mu_{j+1})$ must define a subdiffusion or superdiffusion  on $[\tau_{j+1},\tau_{j+2})$.\newline

\indent In the sequel, $P_{\pmb{\mathfrak{h}},\pmb{\mu},\pmb{\sigma}}^{\pmb{\tau}}$ denotes the measure induced by the stochastic process $(X_t)$ solution of \eqref{eq:sde2}. We define the subscripts $\pmb{\mathfrak{h}},\pmb{\mu},\pmb{\sigma}$ and $\pmb{\tau}$ as follows:
\begin{itemize}
\item $\pmb{\tau}=(\tau_j)_{j=1\ldots N} \in \mathbb{R}^{N+*}$ is the vector of change points with $\tau_1<\tau_2<\dots<\tau_N$,
\item $\pmb{\mathfrak{h}}=(\mathfrak{h}_j)_{j=0\ldots N} \in (0, 1)^{N+1}$ is the vector of Hurst index,
\item $\pmb{\mu}=(\mu_j)_{j=0\ldots N}$ is the set of $N+1$ drift functions from $\mathbb{R}^+\times \mathbb{R}^d \to \mathbb{R}^d$,
\item $\pmb{\sigma}=(\sigma_j)_{j-0\ldots N} \in \mathbb{R}^{(N+1)+*}$ is the vector of diffusion coefficients.
\end{itemize}
We set $(\pmb{\mathfrak{h}},\pmb{\mu}, \pmb{\sigma},\pmb{\tau})=(\mathfrak{h},\mu,\sigma,\emptyset)$ if there is no change point.\newline
\indent Finally, we suppose that for each $\tau_j$ there exists $0\leq j^{\star} \leq n$ such that $\tau_j=t_{j^*}$. It means that the change of motion occurs precisely at a sampling time. Then, we define the subtrajectory $\mathbb{X}^j_{n_j}=(X_{\tau_j},\dots,X_{\tau_{j+1}})$ of size $n_j$ generated with diffusion parameters $(\mathfrak{h}_j,\mu_j,\sigma_j)$.\newline
\indent We present a sequential procedure to estimate both the number of change points $N$ and  the vector of change points $(\tau_1,\dots,\tau_N)$. In the next section, we present the sequential procedure as a statistical test.

\section{Change Point Problem  as a Statistical Test}
\label{sec:test}
The basic ingredient of the proposed algorithm is the test statistic developed by \cite{briane2016adaptive}. This statistic allows to classify a trajectory into three groups of diffusion namely Brownian motion subdiffusion and superdiffusion. We compute this statistic on subtrajectories of the analysed trajectory to detect the change points corresponding to a switch of diffusion. In this section, we present briefly the statistic of \cite{briane2016adaptive}, then we define the change point problem as a statistical test. 

%

\subsection{Trajectory Classification with a Three-Decision Test}
\label{subsec:test}

In this subsection, we suppose there is no change point. Consequently, we observe the trajectory $\mathbb{X}_n$ solution of the stochastic differential equation \eqref{eq:sde2} with $\pmb{\tau}=\emptyset$. The objective is to classify $\mathbb{X}_n$ into one of the three types of diffusion namely subdiffusion, superdiffusion or Brownian motion. This classification problem is tackled thanks to a three-decision test by \cite{briane2016adaptive}. The null hypothesis of their test is $H_{0}$ "$ \mathbb{X}_{n}$ is Brownian". The two alternative hypothesises are $H_{1}$ "$\mathbb{X}_{n}$ is subdiffusive" and $H_{2}$ "$\mathbb{X}_{n}$ is superdiffusive".
\cite{briane2016adaptive} use the following test statistic to carry their statistical test:
\begin{equation}
T_n=\frac{\max_{i=1,\dots,n}\left\| X_{t_i}-X_{t_0} \right\|_2}{\sqrt{(t_n-t_0)\hat{\sigma}^2_n}}
\label{eq:test_stat}
\end{equation}
where $\hat{\sigma}_{n}^2=(1/(2n\Delta))\sum\limits_{j=1}^n\|X_{t_j}-X_{t_{j-1}}\|_2^2$ is a consistent estimate of the diffusion coefficient $\sigma$. The statistic of the maximum is scaled to have a standardized measure. Then, under the hypothesis that $\mathbb{X}_n$ is generated by a Brownian motion of diffusion coefficient $\sigma$, $T_n$ does not depend on $\sigma$ nor $\Delta$ but just on the trajectory size $n$ (see \cite{briane2016adaptive}). If $T_n$ is low, it means the process stays close to its initial position during the period $\left[t_0,t_n\right]$ then its is likely that it is a subdiffusion. 
On the contrary, if $T_n$ is large, it means the process goes far away from its starting point, as a superdiffusion does with high probability. Then \cite{briane2016adaptive} define two thresholds $q_1<q_2$ and state that $\mathbb{X}_n$ is subdiffusive if $T_n<q_1$, superdiffusive if $T_n>q_2$ and Brownian otherwise. The type I error of this three-decision test is controlled at level $\alpha$.

\subsection{Global Null and Alternative Hypothesis of the Test}
We adapt the sequential procedure proposed in \cite{cao2015changepoint} to our problem. \cite{cao2015changepoint} observe the outcome of $n$ statistical tests $H_{0j}: p_j \sim \mathcal{U}(0,1)$ against $H_{1j}:p_j \nsim \mathcal{U}(0,1)$ where $p_j$ is the \Pv of the $j^{\text{th}}$ test, $j=1,\dots,n$.
Hypothesis $H_{0j}$ corresponds to observation of noise at location $j$ while $H_{1j}$ matches with the observation of a true signal. The authors assume that there exist clusters of noise ($H_{0j}$) and clusters of true signal ($H_{1j}$). The objective is to detect these clusters. Their global null hypothesis is that $(p_1,\dots,p_n) \sim \mathcal{U}(0,1)$, matching with the situation where there is only noise. Their alternative hypothesis is that clusters of noise alternate with clusters of true signal.

In our settings, we have a sequence of statistical tests $H_{0j}$ "$ \mathbb{X}^{j}_{n_j}$ is Brownian" against $H_{1j}$ "$\mathbb{X}^{j}_{n_j}$ is subdiffusive" or $H_{2j}$ "$\mathbb{X}^{j}_{n_j}$ is superdiffusive" where  $\mathbb{X}^{j}_{n_j}$ are the subtrajectories of the analysed trajectory $\mathbb{X}_n$. Our global null hypothesis is that $\mathbb{X}_n$ is Brownian on $[t_0, t_n]$:
\begin{equation}
H_0:\mathbb{X}_n \text{ is generated from } (\sigma B_t)_{t_0\leq t\leq t_n}.
\label{eq:H0}
\end{equation}
Our alternative hypothesis is that there exist $\tau_0=t_0<\tau_1,\dots,\tau_N<\tau_{N+1}=t_n$ such that:
\begin{enumerate}
\item $\mathbb{X}_n=(\mathbb{X}^{1}_{n_1},\dots,\mathbb{X}^{N}_{n_N})$ where the subtrajectory $\mathbb{X}^{j}_{n_j}$ is generated with diffusion parameters $(\mathfrak{h}_j,\mu_j,\sigma_j)$.
\item For all $j=1,\dots,N$ diffusion parameters $(\mathfrak{h}_j,\mu_j,\sigma_j)$ and $(\mathfrak{h}_{j+1},\mu_{j+1},\sigma_{j+1})$ are associated to different types of diffusion (Brownian, subdiffusion or superdiffusion).
\end{enumerate} 
\begin{rem}
The case where the whole trajectory is subdiffusive or superdiffusive belongs to the alternative hypothesis. In this case there is no change point ($\pmb{\tau}=\emptyset$).
\end{rem}

In the next section, we present the sequential procedure. The parameters of this algorithm can be chosen such that we control the type I error of the aforementioned test at level $\alpha$. In other words, with appropriate parameters, if the trajectory is fully Brownian, we will not detect any change point with probability $1-\alpha$.

\section{Procedure}
\label{sec:proc}
Our procedure comprises three main steps:
\begin{enumerate}
\item detect the potential change points,
\item gather these potential change points in clusters; one cluster is assumed to contain a single change point,
\item estimate the change point in each cluster.
\end{enumerate}
The critical parameter of our method is the size of the local window $k$.
 There are two parameters to detect the potential change point $(\gamma_1,\gamma_2)$ and two parameters defining the clusters $(c,c^\star)$. We explain each step of our procedure in the next subsections.

\subsection{Detecting the Potential Change Points}
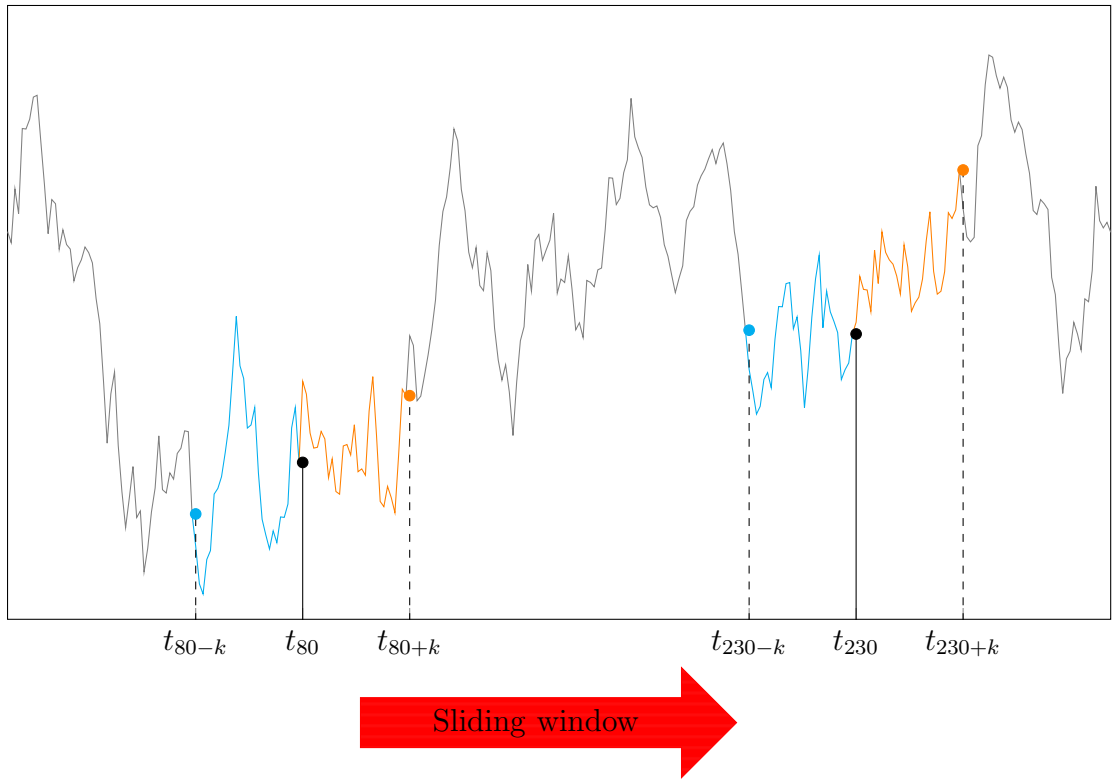
\begin{figure}[ht!]
\centering
\begin{tabular}{c}
\pgfplotstableread {fig2/traj_1dim_ex.dat} {\loadedtable}
\pgfplotsset{compat=newest}
\begin{tikzpicture}[baseline]
\begin{axis}
[xlabel=,ylabel=,
xtick={51,80,109,201,230,259},
xticklabels={$t_{80-k}$,$t_{80}$,$t_{80+k}$,
$t_{230-k}$,$t_{230}$,$t_{230+k}$},
xtick pos=left,
ytick=\empty,
enlarge x limits=false,
ymax=7,
ymin=-12,
height=0.6\linewidth,
width=\linewidth
] 
\addplot[color=gray] table[x=time,y=x1] from \loadedtable;
\addplot[color=cyan] table[x=time,y=x2] from \loadedtable;
\addplot[color=orange] table[x=time,y=x3] from \loadedtable;
\addplot[color=gray] table[x=time,y=x4] from \loadedtable;
\addplot[color=cyan] table[x=time,y=x5] from \loadedtable;
\addplot[color=orange] table[x=time,y=x6] from \loadedtable;
\addplot[color=gray] table[x=time,y=x7] from \loadedtable;

\addplot[dashed] coordinates {
    (51,-15) (51,-8.7396)
};
\addplot[] coordinates {
    (80,-15) (80,-7.1443)
};
\addplot[dashed] coordinates {
    (109,-15) (109,-5.0783)
};

\addplot[dashed] coordinates {
    (201,-15) (201,-3.0506)
};
\addplot[] coordinates {
    (230,-15) (230,-3.1688)
};
\addplot[dashed] coordinates {
    (259,-15) (259,1.9077)
};

\addplot[mark=*,color=cyan] coordinates {
(51,-8.7396)
};
\addplot[mark=*] coordinates {
(80,-7.1443)
};
\addplot[mark=*,color=orange] coordinates {
(109,-5.0783)
};

\addplot[mark=*,color=cyan] coordinates {
(201,-3.0506)
};
\addplot[mark=*] coordinates {
(230,-3.1688)
};
\addplot[mark=*,color=orange] coordinates {
(259,1.9077)
};

\end{axis}
\end{tikzpicture}\\
\tikzfancyarrow[5cm]{Sliding window}\\
\end{tabular}
\caption[Illustration of the sequential procedure on a one dimension toy trajectory]{Illustration of the sequential procedure on a one dimension toy trajectory. \textcolor{cyan}{Blue} parts are the \textcolor{cyan}{backward subtrajectories} on which we compute $B_i$. \textcolor{orange}{Orange} parts are the \textcolor{orange}{forward subtrajectories} on which we compute $A_i$. The black points are the centres of the \textcolor{cyan}{backward} and \textcolor{orange}{forward} subtrajectories. We shift the \textcolor{cyan}{backward} and \textcolor{orange}{forward} subtrajectories all along the trajectory to compute the sequence of ($B_i,A_i$), as shown by the red arrow. }
\label{fig:seq_proc}
\end{figure}

\begin{figure}[ht!]
\centering
\pgfplotstableread {fig2/traj_cp_cut_ex.dat} {\loadedtable}
\pgfplotsset{compat=newest}
\begin{tikzpicture}[baseline]
\begin{axis}
[xlabel=,ylabel=,
minor tick num=4,
xtick=\empty,ytick=\empty,
axis equal,
unit rescale keep size=false,
ymax=15,
ymin=-5,
width=\linewidth
] 
\addplot[color=gray] table[x=x_beg,y=y_beg] from \loadedtable;
\addplot[color=gray] table[x=x_end,y=y_end] from \loadedtable;
\addplot[color=cyan] table[x=x_left,y=y_left] from \loadedtable;
\addplot[color=orange] table[x=x_right,y=y_right] from \loadedtable;
\draw[color=green] (axis cs:-6.5314,5.8334) circle [radius=4.0063];
\draw[color=red] (axis cs:-6.5314,5.8334) circle [radius=17.6373];
\draw[fill=black] (axis cs:-6.5314,5.8334) circle [radius=0.2];
\end{axis}
\end{tikzpicture}
\caption[Illustration of the detection step on a simulated trajectory]{Illustration of the detection step on a simulated trajectory. The trajectory is a zoom of the trajectory represented in Figure \ref{fig:traj_cp} (a) on the portion corresponding to the time interval $[50 ,140]$. First the trajectory is Brownian, then after $\tau_1=100$ it undergoes \BRD. The change point  {$\tau_1$} is represented by the black dot. The \textcolor{cyan}{blue} (respectively \textcolor{orange}{orange}) part corresponds to the subtrajectory \textcolor{cyan}{$\mathbb{X}^-_{\tau_1}$} respectively \textcolor{orange}{$\mathbb{X}^+_{\tau_1}$}. The size of the subtrajectories  is $k=30$. The \textcolor{gray}{gray} part is the complementary part of the trajectory. The two circles are centred on $X_{\tau_1}$. They represent the thresholds defining the different kinds of motion. The radius of the inner (respectively outer) circle is $\gamma_1 \sigma \sqrt{k \Delta}$ (respectively $\gamma_2 \sigma \sqrt{k \Delta}$). For the didactic purpose of the illustration, here we consider $\sigma$ known (we estimate it in our method). The \textcolor{cyan}{blue subtrajectory} stays inside the inner circle: it is classified as subdiffusive. The \textcolor{orange}{orange subtrajectory} goes outside the outer circle: it is  classified as superdiffusive. If the maximum of one part had lied between the two circles it would have been classified as Brownian. The maximum of the \textcolor{cyan}{blue} and \textcolor{orange}{orange} subtrajectories lie in different regions defined by the two limit circles, then $\tau_1=100$ is detected as a potential change point.}
\label{fig:ex}
\end{figure}
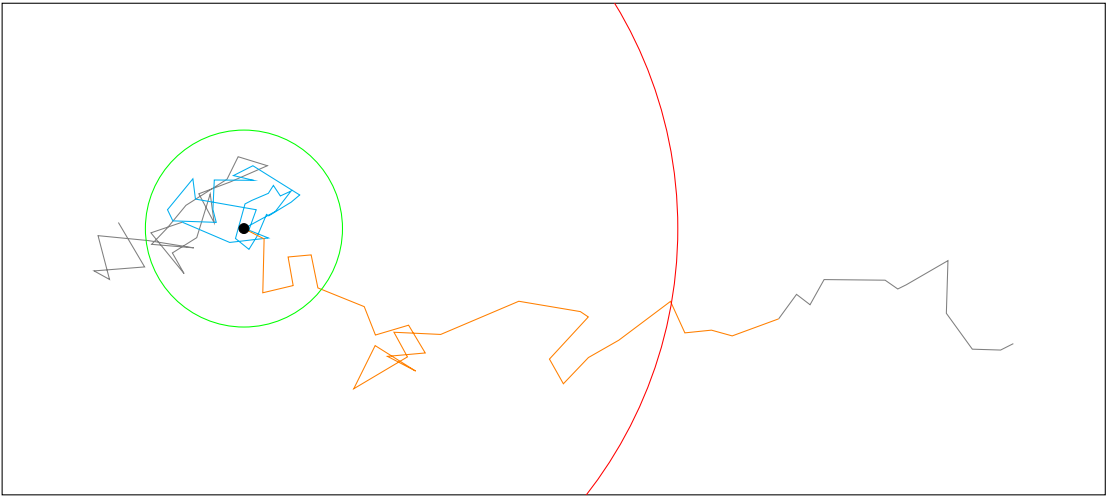

Let $1\leq k\leq n/2$. We will discuss about the choice of $k$ in Section \ref{sec:algo_sim}. For all index $i$ such that $t_k\leq t_i\leq t_{n-k},$ we consider two 
subtrajectories of size $k$ starting at $X_{t_i}$,
\begin{itemize}
\item the backward trajectory $\mathbb{X}_i^{-}=\{X_{t_i},X_{t_{i-1}},\ldots X_{t_{i-
k}}\},$
\item the forward trajectory $\mathbb{X}_i^{+}=\{X_{t_i},X_{t_{i+1}},\ldots 
X_{t_{i+k}}\}.$
\end{itemize}
We compute the test statistic \eqref{eq:test_stat} for the backward and forward trajectory as,
\begin{equation}
B_i=\frac{\max_{j=1,\dots,k}\left\| X_{t_{i-j}}-X_{t_i} \right\|}{\sqrt{(t_{i+k}-t_i)}\hat{\sigma}(t_{i-k}:t_{i})}, \qquad A_i=\frac{\max_{j=1,\dots,k}\left\| X_{t_{i+j}}-X_{t_i} \right\|}{\sqrt{(t_{i+k}-t_i)}\hat{\sigma}(t_i:t_{i+k})}. 
\label{eq:stat_ba}
\end{equation}
where $\hat{\sigma}(t_i:t_{i+k})$ (respectively $\hat{\sigma}(t_{i-k}:t_{i})$) denotes the estimate of the diffusion coefficient from the forward trajectory $\mathbb{X}^+_i$ (respectively the backward trajectory $\mathbb{X}^-_i$). We note that if we use the estimate of the diffusion coefficient proposed in Section \ref{eq:test_stat}, we have $\hat{\sigma}(t_{i-k}:t_i)=\hat{\sigma}(t_{i}:t_{i-k})$.
The denomination $B_i$ (respectively $A_i$) is for \textit{"Before time $t_i$"} (respectively \textit{"After time $t_i$"}). We illustrate this sequential procedure on Figure \ref{fig:seq_proc}.\newline
\newline
\indent Now, we want to compare the backward statistic $B_i$ and the forward statistic 
$A_i.$ The principle is that if the two values are in the same range of values, it is unlikely that
time $t_i$ is a change point. Then, we use two cut-off 
values $\gamma_1<\gamma_2$ which depend on the parameters of the 
procedure, and we define the following step-function:
\begin{equation}
\phi(x;\gamma_1,\gamma_2)=
\left\{
\begin{array}{ll}
1 & \text{if } x<\gamma_1\\
2 & \text{if } x>\gamma_2\\
0 & \text{otherwise.}
\end{array}
\right.
\end{equation}
We have the following interpretation of the cut-off values: $\phi(A_i;\gamma_1,\gamma_2)=0$ means that $\mathbb{X}^+_i$ is Brownian, $\phi(A_i;\gamma_1,\gamma_2)=1$ means that $\mathbb{X}^+_i$ is subdiffusive and $\phi(A_i;\gamma_1,\gamma_2)=2$ superdiffusive.
Then we compute:
\begin{equation}
Q_i=\phi(A_i;\gamma_1,\gamma_2)-\phi(B_i;\gamma_1,\gamma_2), \quad i=k,\dots,n-k.
\end{equation}
If $Q_i=0$ it means that the statistics $B_i$ and $A_i$ belong to the same range of values defined by $\gamma_1$ and $\gamma_2$. Then, both $\mathbb{X}^+_i$ and $\mathbb{X}^-_i$ are  from the same type of diffusion: it is unlikely that $t_i$ is a change point. On the contrary, if $Q_i\neq 0$ the subtrajectories $\mathbb{X}^+_i$ and $\mathbb{X}^-_i$ are not from the same type of diffusion and $t_i$ is a potential change point. The detection step is illustrated on a simulated trajectory in Figure \ref{fig:ex}.\newline

\subsection{Gathering the Potential Change Points into Clusters} 
A first option proposed by \cite{cao2015changepoint} is to consider that a cluster is composed of successive index $i$ (in their context location in DNA sequence) such that $Q_i\neq 0$. \cite{cao2015changepoint} require the cluster to have a minimal size $r^{\star}$ set to $k/2$. However, in our case, this choice does not work well. Due to the high level of randomness of the stochastic processes modelling the trajectory, we observe rarely clusters of size $k/2$ of successive position $i$ in the trajectory such that $Q_i\neq 0$. Then the procedure does not detect any change point (low power of the test). Also, optimizing the minimal size $r^{\star}$ is tricky and can lead to overdetection or underdetection depending on the situation. Therefore we choose an other way to build clusters. Even if it is hard to observe successive potential change points, we argue that a subset of indexes where the concentration of potential change point is high (even if there are not connected) is likely to contain a true change point. Then, we define a cluster of potential change points as a subset of index $\mathcal{M}=\{i,\dots,i+l\}$ such that:
\begin{equation}
\sum\limits_{j=m}^{m+c-1} \mathbf{1}(Q_j\neq 0)\geq c^\star, \quad \forall m=i,\dots,i+l-c+1,
\label{eq:clusters}
\end{equation}
where $c$ and $c^{\star}$ are tuning parameters. We set $c=k/2$, therefore the cluster has a minimal size of $k/2$ as in \citep{cao2015changepoint}. A cluster is created if there are at least $c^\star$ potential change points in a set of $c$ successive points. The parameter $c^\star$ defines the minimum concentration of potential change point needed to build a cluster. Intuitively, we should have $c^{\star}\geq c/2$: the concentration of potential change points $i$ ($Q_i \neq 0$) is higher than the concentration of points $i$ such that $Q_i=0$. We set $c^\star=0.75c$. We note that some points of the clusters are not potential change points ($Q_i=0$).  We emphasize that the choice $c^\star=c$ is equivalent to build clusters as presented in \citep{cao2015changepoint}.\newline 
\indent To illustrate the construction of the clusters, we reproduce a portion of the sequence of $\mathbf{1}(Q_j\neq 0)$ computed on a trajectory simulated with the same parameters as the trajectory presented in Figure \ref{fig:traj_cp} (b):\newline 

\noindent\texttt{0     0     0     1     0     0     
\textcolor{blue}{
1     0     0     1     0     1     0     1     1     0     1     1     1     1     1     1     1     0     0     0     1     1     1     1     1     1     1     1     1     0     1     1     1     0     1     1}
0     0     0     0     0     0     0     1     0     0     0     0     0     0     0     0     0     0     0     0     0     0}.\\

\noindent We use a window size of $k=30$. Then the default parameters of the clustering step are $c=15$ and $c^{\star}=10$. The blue part defines a cluster. In fact, along this cluster, the condition \eqref{eq:clusters} is respected. It turns out that the index of a true change point is contained in this cluster. We would not have detected this change point with the choice of \cite{cao2015changepoint} (corresponding to $c^\star=c$). To detect the same cluster with the choice of \cite{cao2015changepoint}, we should have observed the following sequence of $\mathbf{1}(Q_j\neq 0)$:\newline

\noindent\texttt{0     0     0     1     0     0     
\textcolor{blue}{
1     1     1     1     1     1     1     1     1     1     1     1     1     1     1     1     1     1     1     1     1     1     1     1     1     1     1     1     1     1     1     1     1     1     1     1}
0     0     0     0     0     0     0     1     0     0     0     0     0     0     0     0     0     0     0     0     0     0}.\\

\subsection{Estimating the Change Point in each Cluster}
Denote $\mathcal{M}_k$ the $k^{\text{th}}$ cluster. We estimate the change point of cluster $\mathcal{M}_j$ by:
\begin{equation}
\hat{\tau}_j=t_{r_j}, \quad r_j=\max_{i\in \mathcal{M}_j}|B_i-A_i|.
\end{equation}
We choose the point $i$ of the cluster for which $B_i$ and $A_i$ are the most different. The rational of this idea is that, at the exact change point $\tau_j=t_{r_j}$, $\mathbb{X}^-_{r_j}$ and $\mathbb{X}^+_{r_j}$ are trajectories generated from different diffusion processes and thus $B_{r_j}$ and $A_{r_j}$ must be the most different. At a point $t_i$ close to $\tau_j$, the subtrajectories  $\mathbb{X}^-_i$ and $\mathbb{X}^+_i$ are composed of a mixture of diffusion. Then $B_i$ and $A_i$ reflect this mixture and $|B_i-A_i|\leq|B_{r_j}-A_{r_j}|$.\newline
\indent Finally we can summarize the method as follows:
\begin{proc}
\quad 
\begin{enumerate}
\item For a chosen window size $k$ compute $B_i$ and $A_i$ in \eqref{eq:stat_ba} for $i=k,\dots,n-k$.
\item For prespecified cut-off values $\gamma_1<\gamma_2$ compute $Q_i=\phi(A_i;\gamma_1,\gamma_2)-\phi(B_i;\gamma_1,\gamma_2)$.
\item Decompose $\{k,\dots,n-k\}=W_0\cup W_1$ where $i \in W_0$ if $Q_i=0$ and $i \in W_1$ if $Q_i\neq 0$..
\item Gather the potential change points, that is points $t_i$ such that $Q_i\neq 0$, into clusters   $\mathcal{M}_1,\dots,\mathcal{M}_{\hat{N}} $ satisfying Equation \eqref{eq:clusters}.
\item For each $\mathcal{M}_j$ let $r_j=\max_{i\in \mathcal{M}_j}|B_i-A_i|$ then $\hat{\tau}_j=t_{r_j}$.
\end{enumerate}
\label{proc_cp}
\end{proc}

The parameters of Procedure \ref{proc_cp} are the size of the window $k$, the parameters defining the clusters $c$ and $c^\star$ and the cut-off-values $(\gamma_1,\gamma_2)$. We recommend to set $c=k/2$ and $c^\star=0.75 c$. A choice for the cut-off values $(\gamma_1,\gamma_2)$ is given in Section \ref{sec:cut_off}. Then, the only free parameter to be set by the user is the window size $k$. The influence of parameter $k$ is discussed in Section \ref{sec:algo_sim}.

\section{Cut-off Values}
\label{sec:cut_off}
We choose $\gamma_1$ and $\gamma_2$ such that we control the type I error at level $0<\alpha<1$ that is:
\begin{equation}
P_{1/2,0,\sigma}^\emptyset(\exists i \in \{k,\dots,n^{\star}\}, \quad \sum\limits_{j=i}^{i+c-1}\mathbf{1}(Q_j\neq0)\geq c^{\star})\leq \alpha,
\label{eq:prob_typeI}
\end{equation}
where $n^{\star}=n-k-c+1$. We explain why controlling the probability in \eqref{eq:prob_typeI} at level $\alpha$ is equivalent to control the type I error at level $\alpha$. The left hand side of Equation \eqref{eq:prob_typeI} is the probability to build one cluster of minimal size $c$ (in the sense of \eqref{eq:clusters}) under $H_0$. With Procedure \ref{proc_cp}, we need to build a cluster of potential change points to detect a change point, otherwise no change point is detected. Then, controlling the probability in \eqref{eq:prob_typeI} at level $\alpha$ under $H_0$ is equivalent to control the probability to detect falsely a change point under $H_0$ at level $\alpha$ (definition of the type I error). Now we have the following proposition:

\begin{prop}\label{prop:gamma12}
Let define $d_i=\min(B_i,A_i)$ and $D_i=\max(B_i,A_i)$ where $A_i$ and $B_i$ are the test statistics \eqref{eq:stat_ba}, for $i=k,\dots,n^{\star}$. We define $\gamma_1^\star$ and $\gamma_2^\star$ as:
\begin{equation}
\begin{split}
P_{1/2,0,\sigma}^\emptyset\Bigl(\min_{i=k,\dots,n^\star} d_{i(c^\star/2)}<\gamma_1^\star\Bigr)&=\frac{\alpha}{2},\\
P_{1/2,0,\sigma}^\emptyset\Bigl( \max_{i=k,\dots,n^\star} D_{i(c-c^\star/2)}>\gamma_2^\star\Bigr)&=\frac{\alpha}{2},
\end{split}
\label{eq:gamma_opt}
\end{equation}
where $d_{i(c^\star/2)}$ is the $c^\star/2$ smallest element of $(d_i,\dots,d_{i+c-1})$ and $D_{i(c-c^\star/2)}$ the $c-c^\star/2$ smallest element of $(D_i,\dots,D_{i+c-1})$ (equivalently the $c^\star/2$ greatest element).
In other words, $\gamma_1^\star$ is the quantile of order $\alpha/2$ of the random variable $\min_{i=k,\dots,n^\star} d_{i(c^\star/2)}$ and $\gamma_2^\star$ is the quantile of order $1-\alpha/2$ of the random variable $\max_{i=k,\dots,n^\star} D_{i(c-c^\star/2)}$. With the choice of cut-off values $\gamma_1^\star$ and $\gamma_2^\star$ Procedure \ref{proc_cp} with parameters $(k,c,c^\star)$ controls the type I error \eqref{eq:prob_typeI} at level $\alpha$.

\end{prop}
A proof of Proposition \ref{prop:gamma12} is available in the Supplementary Materials.We estimate $\gamma_1^\star$ and $\gamma_2^\star$ with Monte-Carlo simulation, see the Monte Carlo Algorithm in the Supplementary Materials. \newline
\indent When we analyse the proof of Proposition \ref{prop:gamma12} (Supplementary Materials), we realize that the choice $(\gamma_1^\star,\gamma_2^\star)$ is not optimal. In particular, the bound in Equation 
(1.7) (in Supplementary Materials) is loose. Then, we can see from simulations that the probability of type I error is controlled at a much lower level than $\alpha$ (about 1\% when $\alpha=5\%$ see Table \ref{tab:typeI_gamma12}). Consequently, we recommend to use the cut-off values verifying:
\begin{equation}
\begin{split}
P_{1/2,0,\sigma}^\emptyset\Bigl(\min_{i=k,\dots,n^\star} d_{i(c^\star)}<\tilde{\gamma_1}\Bigr)&=\frac{\alpha}{2},\\
P_{1/2,0,\sigma}^\emptyset\Bigl( \max_{i=k,\dots,n^\star} D_{i(c-c^\star)}>\tilde{\gamma_2}\Bigr)&=\frac{\alpha}{2}.
\end{split}
\label{eq:gamma_opt2}
\end{equation}

We replace $c^{\star}/2$ in Equation \eqref{eq:gamma_opt} by $c^{\star}$. Then, it is straightforward to show that $\gamma_1^\star\leq \tilde{\gamma}_1$ and $\gamma_2^\star  \geq \tilde{\gamma}_2$. 
We deduce from these inequalities that the power of Procedure \ref{proc_cp}, that is its ability to find a true change point, is higher with the choice $(\tilde{\gamma}_1,\tilde{\gamma}_2)$ than with ($\gamma_1^\star,\gamma_2^\star$). In fact, a high value of $\gamma_1$ detects better subdiffusions than a low value of $\gamma_1$. The other way around, a low value of $\gamma_2$ detects better superdiffusions than a high value of $\gamma_2$. As stated before, the control the type I error constrains the choice of $(\gamma_1,\gamma_2)$.  We show from simulations that the choice $(\tilde{\gamma}_1,\tilde{\gamma}_2)$ still controls the type I error at level $\alpha$ (see Table \ref{tab:typeI_gamma12}). Estimations of ($\gamma_1^\star,\gamma_2^\star$) and $(\tilde{\gamma}_1,\tilde{\gamma}_2)$ --obtained with the Monte Carlo algorithm, see Supplementary Materials-- are given in Table \ref{tab:gamma12}.

\begin{table}[ht!]  
\small                             
\centering     
\caption[Control of the type I error for different cut-off values $(\gamma_1,\gamma_2)$.]{Control of the type I error for different cut-off values $(\gamma_1,\gamma_2)$. We estimate the probability of type I error with the proportion of trajectories with at least one change point detected among $100\,001$ Brownian trajectories. We use the default parameters for Procedure \ref{proc_cp} that is $c=k/2$ and $c^{\star}=0.75 c$. The estimations are accurate at $\pm 0.14 \%$.}                      
\label{tab:typeI_gamma12}                                
\begin{tabular}{cccc}  
\hline
 & & \multicolumn{2}{c}{\textbf{Probability of Type I error}} \\
 \cline{3-4}
$n $  & $k$ & with $(\gamma_1^{\star},\gamma_2^{\star})$  &  with $(\tilde{\gamma}_1,\tilde{\gamma}_2)$\\ \hline                  
150 & 20 &  0.60 & 5.21  \\
150 & 30 &  0.65 & 4.81  \\
150 & 40 &  0.94 & 4.56  \\
300 & 20 &  0.47 & 5.04  \\
300 & 30 &  0.59 & 4.89  \\
300 & 40 &  0.82 & 4.83  \\
\hline 
\end{tabular}                                                   
\end{table}

\begin{table}[ht!]  
\small                              
\centering     
\caption[Cut-off values of Procedure \ref{proc_cp}]{Cut-off values of Procedure \ref{proc_cp} for different sizes of trajectory $n$ and sizes of window $k$. The cut-off values are estimated with the Monte Carlo Algorithm in Supplementary Materials using $N=10\,001$ replications.}                      
\label{tab:gamma12}                                 
\begin{tabular}{ccccccc}  
\hline
$n $  & $k$ &$\gamma_1^{\star}$  & $\gamma_2^{\star}$ & $\tilde{\gamma}_1$ & $\tilde{\gamma}_2$\\ \hline                  
150 & 20 &  0.61 & 3.38 & 0.74 & 3.09 \\
150 & 30 &  0.65 & 3.35 & 0.78 & 3.05 \\
150 & 40 &  0.68 & 3.28 & 0.80 & 3.03 \\
300 & 20 &  0.58 & 3.55 & 0.71 & 3.27 \\
300 & 30 &  0.62 & 3.55 & 0.74 & 3.26 \\
300 & 40 &  0.64 & 3.52 & 0.75 & 3.25 \\
\hline 
\end{tabular}                                                   
\end{table}


\section{Performance of the Method on Simulations}
\label{sec:algo_sim}
%


We simulate two different scenarios, see Table \ref{tab:scenario}. We limit this study to the two-dimensional case $d=2$.
\subsection{Simulation Settings}
 Subdiffusions are modelled by Ornstein-Uhlenbeck process:
\begin{equation}
dX^i_t=-\lambda (X^i_t-\theta_i)dt+ \sigma dB^{1/2,i}_t,\qquad i=1,2,
\label{eq:sde_OU2}
\end{equation}
where $\lambda>0$ models the restoring force toward the equilibrium point $\theta=(\theta_1,\theta_2)$; $\sigma>0$ is the diffusion coefficient.
For modelling superdiffusion we use the \BRD solution of the SDE:
\begin{equation}
dX^i_t=(v/\sqrt{2})dt+\sigma dB^{1/2,i}_t,\qquad i=1,2,
\label{eq:sde_dir2}
\end{equation}
where $\sigma>0$ is the diffusion coefficient and $v>0$. Then the constant drift $\mathbf{v}=(v,v)/\sqrt{2}$ verifies $\norme{\mathbf{v}}=v$.

For each scenario, we compute the performances of our procedure for different values of the parameters $v$ (for the \BRD) and $\lambda$ (for the \OU process).
We assess the performances of our algorithm with respect to two criteria:
\begin{enumerate}
\item the number of change points detected,
\item the location of these change points. 
\end{enumerate}

Criterion 2 is assessed only on the trajectories for which we detect the right number of change points that is $N=2$. We compute the average and standard deviation of the locations.  We analyse the results of the simulation 
on the different scenarios in the next paragraphs.

\begin{table}[ht!]
\small
\centering
\caption[Simulation scenarios for the Monte Carlo study]{Simulation scenarios for the Monte Carlo study. The size of 
the simulated trajectories is $n = 300.$  The change points occur at $\tau_1 = 
100$ and $\tau_2 = 175.$ We set $\sigma=1$ for the diffusion coefficient and $\Delta=1$ for the step of time. For the \OU process \eqref{eq:sde_OU2}, we define the equilibrium point as $\theta=X_{\tau_1}$ where $X_{\tau_1}$ is the position of the particle at $\tau_1$.}\label{tab:scenario}
\begin{tabular}{lll}
\hline 
Times & Scenario 1 & Scenario 2\\
\hline 
$[1, 100]$ & Brownian  & Brownian\\
$[101, 175]$ & Brownian with drift & \OU \\
$[176, 300]$ & Brownian & Brownian\\
\hline 
\end{tabular} 
\end{table}

\subsection{Scenario 1}

First, we illustrate the scenario of simulation with Figure \ref{fig:traj_cp} (a) showing a trajectory simulated with Scenario 1.
Table \ref{tab:brd} gives us the results associated to Scenario 1 (see Table \ref{tab:scenario}).  We can see clearly that, as $\norme{v}$ increases, the performance of the method increases with respect to both criteria. For a given window size $k$, we get:
\begin{enumerate}
\item the proportion of trajectories for which we detect the right number of change point ($\hat{N}-N=0$) tends to 1 as $v$ inceases.
\item given $\hat{N}-N=0$, the bias and the variance of the estimated change point decrease to $0$  as $v$ inceases.
\end{enumerate}

We also notice that for the window size $k=20$, the performance of the algorithm is lower than for $k=30$ and $40$ except when $\norme{v}=2$. As the size of the window is too low, it is hard for the algorithm to detect a \BRD with a low drift norm. In particular, when $\norme{v}=0.6$, it does not detect any change point in most cases; we note that $\hat{N}-N=-2$ for 42.2\% of the trajectories. However, when the drift norm is high, a low window size performs as  good as the larger ones (see the case $\norme{v}=2$). It performs even better if the change points $\tau_1,\tau_2$ are closer. In this case, a large window tends to mix up the two change points and consequently find only one.  We can summarize this as follows: a large window size enables to detect well the change points associated with a small drift $v$ if the change points are significantly separated while a small window is able to distinguish two close change points if the drift $v$ is large enough. \newline
\begin{table}[ht!]
\centering
\caption[Performance of the Procedure \ref{proc_cp} for Scenario 1]{Performance of the Procedure \ref{proc_cp} for Scenario 1 (see \ref{tab:scenario}) for different window sizes $k$ and different values of the drift $v$.
The computations are based on $1\,001$ simulated trajectories from Scenario 1. We compute the proportions of trajectories with $\hat{N}-N=-2,$ $\hat{N}-N=\pm 1,$ 
$\hat{N}-N=0$ and $\hat{N}-N\geq 2.$ The column $\tau_1$ (respectively 
$\tau_2$ ) gives the empirical average of the first (respectively second) detected 
change point on 300 trajectories among which we detect the right number of 
change points ($\hat{N}-N=0$). The number in brackets is the empirical standard deviation of the estimate of $\tau_1$ and $\tau_2$ computed on these $300$ trajectories.
}\label{tab:brd}
\begin{tabular}{cccccccll}
\hline
\rule{0pt}{3ex} & &\multicolumn{5}{c}{$\hat{N}-N$}& &  \\
\cline{3-7}
$v$ & $k$ & -2 & -1 & 0 & 1 & $\geq2$ & \multicolumn{1}{c}{$\tau_1$} & \multicolumn{1}{c}{$\tau_2$} \\
\hline
0.6 & 20 & 42.2 & 14.1 & 34.7 & 5.7 & 3.3 & 126.3 ~ (23.7) & 153.7 ~ (23.6) \\
0.6 & 30 & 20.9 & 16.9 & 55.9 & 5.5 & 0.8 & 115.0 ~ (17.8) & 162.8 ~ (18.4)  \\
0.6 & 40 & 11.8 & 18.6 & 67.6 & 1.8 & 0.2 & 109.4 ~ (15.5) & 168.2 ~ (15.1)  \\
0.8 & 20 & 6.5 & 12.9 & 54.4 & 17.3 & 8.9 & 117.4 ~ (16.7) & 157.5 ~ (18.5) \\
0.8 & 30 & 1.6 & 6.5 & 84.1 & 6.3 & 1.5 & 107.3 ~ (11.6) & 170.1 ~ (14.1) \\
0.8 & 40 & 0.3 & 4.1 & 93.2 & 2.3 & 0.1 & 104.7 ~ (9.7) & 172.4 ~ (10.0) \\
1 & 20 & 0.2 & 3.7 & 63.2 & 21.6 & 11.3 & 108.3 ~ (12.1) & 168.2 ~ (13.2) \\
1 & 30 & 0.0 & 1.9 & 93.8 & 3.5 & 0.8 & 102.9 ~ (5.9) & 173.9 ~ (6.5) \\
1 & 40 & 0.0 & 0.1 & 97.7 & 1.9 & 0.3 & 103.2 ~ (6.6) & 174.5 ~ (7.3) \\
2 & 20 & 0.0 & 0.0 & 96.6 & 2.4 & 1.0 & 101.4 ~ (2.1) & 176.0 ~ (2.3) \\
2 & 30 & 0.0 & 0.3 & 96.8 & 2.5 & 0.4 & 101.2 ~ (3.4) & 175.9 ~ (2.6) \\
2 & 40 & 0.0 & 0.1 & 99.2 & 0.5 & 0.2 & 101.5 ~ (2.8) & 175.8 ~ (2.9) \\
\hline
\end{tabular}


\end{table}

\subsection{Scenario 2}
\begin{figure}[t!]
\centering
\begin{tabular}{cc}
\pgfplotstableread {fig2/traj_cp_ex2.dat} {\loadedtable}

\begin{tikzpicture}[baseline]
\begin{axis}
[xlabel=,ylabel=,
axis equal,
unit rescale keep size=true,
xmax=52,
xmin=-13,
minor tick num=4,
xtick=\empty,ytick=\empty,
enlarge y limits=false,
width=0.56\linewidth,
]
\addplot[color=blue] table[x=x_brd_beg,y=y_brd_beg] from \loadedtable;
\addplot[color=blue] table[x=x_brd_end,y=y_brd_end] from \loadedtable;
\addplot[color=red] table[x=x_brd_mid,y=y_brd_mid] from \loadedtable; 
\draw[fill=yellow] (axis cs:-5.2554,6.5292) circle [radius=5];
\draw[fill=yellow] (axis cs:36.1714,-10.0091) circle [radius=5];
\end{axis}
\end{tikzpicture} &  \pgfplotstableread {fig2/traj_cp_ex2.dat} {\loadedtable}

\begin{tikzpicture}[baseline]
\begin{axis}
[xlabel=,ylabel=,
axis equal,
minor tick num=4,
xtick=\empty,ytick=\empty,
enlarge x limits=false,
enlarge y limits=false,
width=0.56\linewidth,
]
\addplot[color=green] table[x=x_ou_mid,y=y_ou_mid] from \loadedtable; 
\addplot[color=blue] table[x=x_ou_beg,y=y_ou_beg] from \loadedtable;
\addplot[color=blue] table[x=x_ou_end,y=y_ou_end] from \loadedtable;
\draw[fill=yellow] (axis cs:19.6630,-4.1186) circle [radius=2];
\draw[fill=yellow] (axis cs:17.3251,-6.3055) circle [radius=2];
\addplot[mark=*,color=white] coordinates {
(1,11) (1,-13)
};
\end{axis}
\end{tikzpicture}\\
(a) \vspace{0.5cm}  & (b)\\
\end{tabular}
\caption[Simulated trajectories from Scenario 1 and 2]{Simulated trajectories. Figure (a), trajectory from Scenario 1 with $v=0.8$. We detect $\hat{N}=2$ change points $\hat{\tau}_1=99$ and $\hat{\tau}_2=169$ with a window of size 30. Figure (b) trajectory from Scenario 2 with $\lambda=1$. We detect $\hat{N}=2$ change points $\hat{\tau}_1=87$ and $\hat{\tau}_2=165$ with a window of size 30. The locations of the change points $X_{\hat{\tau}_1}$ and $X_{\hat{\tau}_2}$ are shown as yellow dots on the trajectories.}
\label{fig:traj_cp}
\end{figure}
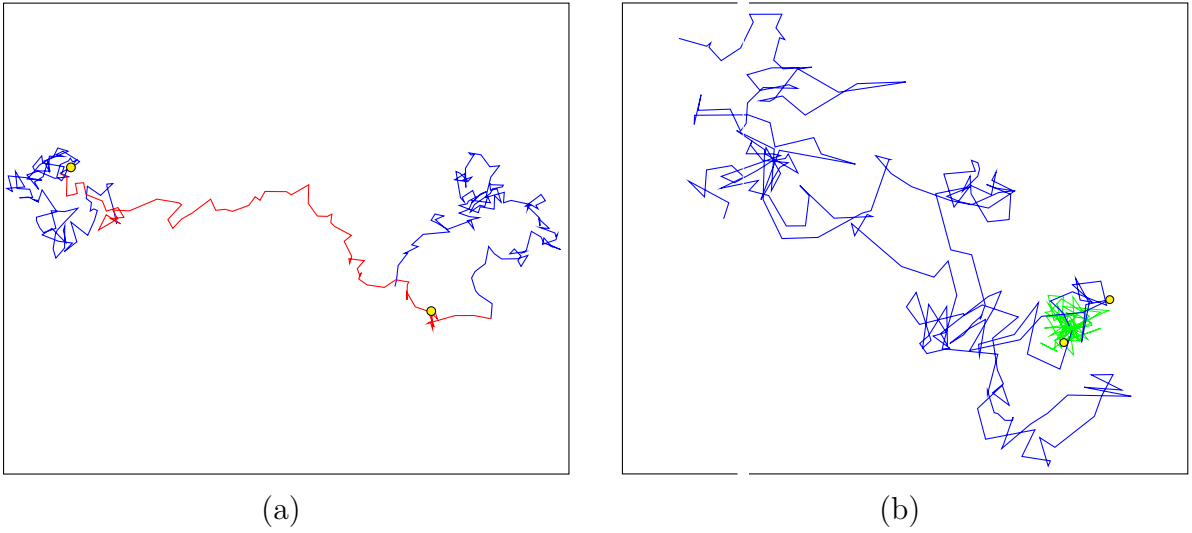

We illustrate the scenario of simulation with Figure \ref{fig:traj_cp} (b) showing a trajectory simulated with Scenario 2.
Table \ref{tab:ou} gives us the results associated to Scenario 2 (see Table \ref{tab:scenario}). As in Scenario 1, for a window size $k=20$ the performance of the algorithm increases as $\lambda$ increases. However, it does not behave the same way if the window size is $30$ or $40$. For $k=30$, the performance increases from $\lambda=1$ to $\lambda=2$ but remains the same for larger values of $\lambda$. For the window size $k=40$, the proportion of trajectories with the correct number of detected change points dramatically drops from 83.6\% with $\lambda=1$ to 54.1\% for $\lambda=4$. At the same time, the proportion of trajectories with $\hat{N}-N=-1$ increases. It means that when $\lambda$ becomes too high the algorithm mixes up the two change points and find only one. As $\lambda$ is high (clear subdiffusion), we detect a potential change point very early in the trajectory: as soon as few points of the forward subtrajectory $\mathbb{X}_i^+$ enter in the subdiffusion regime $(t\geq \tau_1)$  we classify it as subdiffusive. For example, if $\lambda$ is big enough we can suppose that the subtrajectory of size $k$ $\mathbb{X}^+_i=(X_{t_i},\dots X_{\tau_1},X_{\tau_1+1},X_{\tau_1+2})$ will be classified as subdiffusive with only three points in the subdiffusive regime. Then, we get a long sequence of potential change points. But as $k$ is large, the forward subtrajectory has already reached the second change point $\tau_2$. Consequently, it begins to detect potential change points corresponding to the second change point $\tau_2$. As there is a single cluster of potential change points, the algorithm only detects one change point instead of the two expected. From our simulations, we observe that the change point detected is either close to $\tau_1$ or $\tau_2$: it estimated correctly one change point out of the two real change points.\newline
\indent The idea is that, in a way, a large $\lambda$ (a very clear subdiffusion) makes the two change points get closer artificially. Then, a large window can not separate them. We note that, from our simulations, in the case of a change point between \BRD and Brownian motion, we do not observe such a phenomenon (that is a fall of the proportions of trajectories with the right number of detected change points when $v$ increases). However, when the change points are close and the window size $k$ is large compared to the gap between the change points, the performance stops increasing (but does not fall) above a certain value of $v$.

\begin{table}[th!]
\centering
\caption[Performance of the Procedure \ref{proc_cp} for Scenario 2]{Performance of the Procedure \ref{proc_cp} for Scenario 2 (see \ref{tab:scenario}) for different window sizes $k$ and different values of parameter $\lambda$. We use the same protocol as in Table \ref{tab:brd}.}\label{tab:ou}
\begin{tabular}{cccccccll}
\hline
\rule{0pt}{3ex} & &\multicolumn{5}{c}{$\hat{N}-N$}& &  \\
\cline{3-7}
$\lambda$ & $k$ & -2 & -1 & 0 & 1 & $\geq 2$ & \multicolumn{1}{c}{$\tau_1$} & \multicolumn{1}{c}{$\tau_2$}  \\
\hline
\rule{0pt}{3ex}1 & 20 & 18.1 & 44.1 & 31.8 & 5.3 & 0.7 & 109.9 ~ (20.7) & 167.8 ~ (17.9) \\
1 & 30 & 0.8 & 16.3 & 78.3 & 4.2 & 0.4 & 104.9 ~ (8.7) & 169.9 ~ (9.3) \\
1 & 40 & 0.0 & 13.2 & 83.6 & 3.0 & 0.2 & 105.6 ~ (10.8) & 170.4 ~ (11.6) \\
2 & 20 & 3.1 & 22.6 & 68.1 & 5.5 & 0.7 & 106.4 ~ (8.5) & 170.2 ~ (8.1) \\
2 & 30 & 0.1 & 6.5 & 89.4 & 3.4 & 0.6 & 107.5 ~ (8.7) & 169.1 ~ (8.2) \\
2 & 40 & 0.0 & 21.2 & 77.0 & 1.6 & 0.2 & 108.0 ~ (12.7) & 169.1 ~ (12.8) \\
3 & 20 & 1.1 & 17.1 & 74.8 & 5.9 & 1.1 & 106.3 ~ (5.6) & 170.1 ~ (7.9) \\
3 & 30 & 0.0 & 5.7 & 90.2 & 3.2 & 0.9 & 108.7 ~ (8.8) & 167.6 ~ (8.7) \\
3 & 40 & 0.1 & 32.1 & 64.8 & 2.5 & 0.5 & 109.3 ~ (12.9) & 166.4 ~ (13.5) \\
4 & 20 & 0.6 & 12.2 & 79.4 & 6.8 & 1.0 & 107.2 ~ (6.5) & 169.9 ~ (8.4) \\
4 & 30 & 0.0 & 6.5 & 89.7 & 3.1 & 0.7 & 109.6 ~ (9.6) & 166.5 ~ (9.2) \\
4 & 40 & 0.0 & 44.3 & 54.1 & 1.4 & 0.2 & 111.5 ~ (13.3) & 166.0 ~ (13.3) \\
\hline
\end{tabular}
\end{table}

\indent Once the change points are estimated, we can label the type of diffusion on each subtrajectory $\mathbb{X}^j_{n_j}=(X_{\hat{\tau}_j},\dots,X_{\hat{\tau}_{j+1}})$. For a given scenario, value of parameter ($\lambda$ or $v$) and size window $k$, we assessed the labelling of the subtrajectories. Results and details of the evaluation are given in Table \ref{app:tab:prop_label}.

\begin{rem}
It is possible that two consecutive subtrajectories $\mathbb{X}^{j-1}_{n_{j-1}}=(X_{\hat{\tau}_{j-1}},\dots,X_{\hat{\tau}_{j}})$ and $\mathbb{X}^j_{n_j}=(X_{\hat{\tau}_j},\dots,X_{\hat{\tau}_{j+1}})$ are labelled as the same type of diffusion. Then, the \textit{a-posteriori} labelling process questions the fact that $\hat{\tau}_{j}$ is a change point. In fact, we recall that a change point is defined as a time at which the particle switches from one type of diffusion to another, see Section \ref{sec:model_cp}. Then, in this case, we consider that $\hat{\tau}_{j}$ is finally not a change point. We note that the results in Table \ref{tab:brd} and \ref{tab:ou} are given without taking into account the \textit{a-posteriori} labelling process.   
\end{rem}


%

\begin{table}[ht!]
\small
\centering
\caption[Proportions of trajectories (among the trajectories with $\hat{N}=N$) for which subtrajectories are correctly labelled, in scenario 1 and 2]{Proportions of trajectories (among the trajectories with $\hat{N}=N$) for which subtrajectories are correctly labelled, in scenario 1 and 2. The change points are detected and estimated with Procedure \ref{proc_cp}. The subtrajectories are labelled using the three-decision test of \cite{briane2016adaptive} at level $5\%$. Columns 3 et 4 (respectively 4 and 5) correspond to scenario 1 (respectively scenario 2). For example, in scenario 1 with $v=0.6$, when we use a window of size $k=20$, 73.7\% of the trajectories for which we detect $N=2$ change points are labelled as Brownian on $[t_0,\hat{\tau}_1]$, superdiffusive on $[\hat{\tau}_1,\hat{\tau}_2]$ and again Brownian on $[\hat{\tau}_2,t_n]$.}\label{app:tab:prop_label}
\begin{tabular}{ccccc}
\hline
\rule{0pt}{3ex} & \multicolumn{2}{l}{Scenario 1}& \multicolumn{2}{l}{Scenario 2} \\
\hline
\rule{0pt}{3ex} $k$ & $v$ & \% right label & $\lambda$ & \% right label \\
\hline
20 & 0.6 & 73.7 & 1 & 83.0 \\
30 & 0.6 & 82.3 & 1 & 89.3 \\
40 & 0.6 & 86.0 & 1 & 85.0 \\
20 & 0.8 & 74.7 & 2 & 90.7 \\
30 & 0.8 & 87.7 & 2 & 89.0 \\
40 & 0.8 & 88.7 & 2 & 88.7 \\
20 & 1.0 & 82.0 & 3 & 89.3 \\
30 & 1.0 & 86.3 & 3 & 87.3 \\
40 & 1.0 & 88.7 & 3 & 86.7 \\
20 & 2.0 & 89.7 & 4 & 88.7 \\
30 & 2.0 & 90.0 & 4 & 85.0 \\
40 & 2.0 & 90.3 & 4 & 84.7 \\
\hline
\end{tabular}
\end{table}

\section{Comparisons with Competitive Methods}
\label{sec:comparison}
We compare our method to two other methods. The method of \cite{turkcan2013bayesian} detects change points between Brownian motion and parametric models of subdiffusion including the \OU process.
The method of \cite{monnier2015inferring} detects change points between Brownian motion and Brownian motion with drift which is an example of superdiffusion.
We note that none of these methods deal with the three types of diffusion (Brownian motion, subdiffusion and superdiffusion) as we do. In this section, we present the two competitive methods and compare their performances to Procedure \ref{proc_cp} on simulations. At the end of the section, we give a particular emphasis on the speed and stability of the different methods.

\subsection{The Method of \cite{turkcan2013bayesian}}

First the method of \cite{turkcan2013bayesian} is a parametric method. The two parametric models under concern are the Brownian motion and the \OU process (called diffusion in a harmonic potential in \citep{turkcan2013bayesian}). \cite{turkcan2013bayesian} select the model that minimizes the BIC criterion. For detecting change points, the BIC criterion is computed on a sliding window along the trajectory. When the BIC indicates a switch of model and that the new model is confirmed in the next $r$ steps of times, a change is assumed to occur. \newline
\indent We reproduce the simulation described in \cite{turkcan2013bayesian}. We simulate $N=100$ trajectories of size $n=500$. First the trajectory undergoes an \OU process and at time $\tau_1=250$ it switches to a Brownian motion. The two processes share the same diffusion coefficient $\sigma=0.4472$. The parameters of the \OU process \eqref{eq:sde_OU2} is $\lambda=7.3870$. The step of time is $\Delta=0.05$. Results of the two methods are given in Table \ref{tab:comp_masson}. We can see that our method show better results in both the number $\hat{N}$ of detected change points and in the location of the change points. We also emphasize that we do not set $r=3$ as in \citep{turkcan2013bayesian} but we set $r=51$ which corresponds to the size of the window. With $r=3$, the method of \cite{turkcan2013bayesian} detects more than 4 change points in 91\% of the trajectories. Actually, the method is able to detect the change point, if a collection of about $N=50$ trajectories showing the same number of change points at the same location is available. Accordingly, it provides good results in average. However, such a situation is not realistic in practical imaging. In our scenarios, our non-parametric method outperforms the parametric method of \cite{turkcan2013bayesian}.

\begin{table}[ht!]
\small
\centering
\caption[Comparison of Procedure \ref{proc_cp} and the method of \cite{turkcan2013bayesian}]{Comparison of Procedure \ref{proc_cp} and the method of \cite{turkcan2013bayesian} on the simulation of \cite{turkcan2013bayesian}. We recall that the true change point is $\tau_1=250$.}
\label{tab:comp_masson}
\begin{tabular}{cccccl}
\hline
\rule{0pt}{3ex} & \multicolumn{4}{c}{$\hat{N}-N$}&   \\
\cline{2-5}
Method &  -1 & 0 & 1 & $\geq 2$ & \multicolumn{1}{c}{$\tau_1$}  \\
\hline
\rule{0pt}{3ex}Procedure \ref{proc_cp} & 19 & 77 & 3 & 1 & 240.5 ~ (29.4)  \\
Method of \cite{turkcan2013bayesian} & 27 & 59 & 14 & 0 & 176.3 ~ (53.7)  \\
\hline
\end{tabular}
\end{table}

\subsection{The Method of \cite{monnier2015inferring}}

\cite{monnier2015inferring} use two parametric models to fit the displacements of the particle: the Brownian motion and the Brownian motion with drift. We note that the Brownian motion can be seen as a Brownian motion with a null drift. The two models are actually a unique parametric model with parameters $v=(v_1,v_2)$ and $\sigma$ (with $v=(0,0)$ for the Brownian case). Then, \cite{monnier2015inferring} use hidden Markov models to fit the displacements of the particle over time. The hidden states are defined as a set of a drift parameter and diffusion coefficient $S_k={v_k,\sigma_k}$. They estimate both the number of states $K$, the parameters ($v_k,\sigma_k$) and the successive (hidden) states along the trajectories. They also add a constrained $v=0$ for modelling Brownian motion. Model selection is used with a Bayesian criterion to select the best model. If we assume that $K\leq 2$ and also consider the constrained models with $v=0$, the competing models are:
\begin{description}
\item[Model 1] a single state which is the Brownian with parameter $\sigma_1$,
\item[Model 2] a single state which is the Brownian with drift with parameters ($v_1$,$\sigma_1$),
\item[Model 3] two states which are two Brownian with parameter $\sigma_1$ and $\sigma_2$,
\item[Model 4] two states which are one Brownian and one Brownian with drift with respective parameters $\sigma_1$ and $(v_2,\sigma_2)$,
\item[Model 5] two states which are two Brownian with drift with parameters $(v_1,\sigma_1)$ and $(v_2,\sigma_2)$.
\end{description} 

In our experiment, we run the method of \cite{monnier2015inferring} on 100 simulated trajectories from Scenario 1. We assume  $K\leq 2$ that is the competing models are the five models aforementioned. Results are given in Table \ref{tab:comp_natmeth}. When $v=0.6$ or $0.8$, \cite{monnier2015inferring} detect no change point $(\hat{N}-N=-2)$ for a large majority of the trajectories. In this case, the selected model can either have one state (models 1 and 2) or have two states but from the same type of diffusion (models 3 and 5). Actually, when $v=0.6,0.8$, the preferred model is Brownian only (model 1) for most of the trajectories (see Table \ref{app:tab:prop_mod_natmed}). Then the drift is too low to select a model involving Brownian with drift. As expected, the performance of the method of \cite{monnier2015inferring} improves as $v$ increases. The method detects the right number of change points for 96\% of the trajectories when $v=2$. When the method detects at least one change point, it means that the selected model is the model 5. Even when the right model is chosen, it can over-detect the number of change points (that is $\hat{N}-N\geq 1$). We have 9\% of overdetection when $v=1$. When the method detects the right number of change points ($\hat{N}-N=0$), the location of the change points are very close to the true locations. For instance when $v=2$, the average location of the first detected change point is 100 (which is exactly $\tau_1$) and its standard deviation is 1.4. Finally, our non-parametric method detects better the change points when the drift is low ($v\leq 0.8$). The quality of detections are similar when the drift is high enough ($v=2$). For $v=1$, we have a larger proportion of trajectories detected with the right number of change points with our method except when we use a window size of $n=20$ (63.2\% with our method \textit{versus} 68\% with the method of \cite{monnier2015inferring}). The locations of the detected change points among the trajectories with $\hat{N}-N=0$ are slightly more accurate with the method of \cite{monnier2015inferring}. 
\begin{table}
\centering
\caption[Performance of the algorithm of \cite{monnier2015inferring} for Scenario 1]{Performance of the algorithm of \cite{monnier2015inferring} for Scenario 1 (see Table \ref{tab:scenario}). The computations are based on 100 simulated trajectories from Scenario 1. We compute the proportions of trajectories with $\hat{N}-N=-2,$ 
$\hat{N}-N=\pm 1,$ $\hat{N}-N=0$ and $\hat{N}-N\geq 2.$ The column $\tau_1$ (respectively $\tau_2$ ) gives the empirical average of the first (respectively second) detected change point on the trajectories among which we detect the right number of change points ( $\hat{N}-N=0$).  The number in brackets is the empirical standard deviation of the estimates of $\tau_1$ and $\tau_2$. We note that the empirical average and standard deviation  estimate of $\tau_1$ and $\tau_2$ are not computed over the same number of trajectories for the different values of the drift $v$ (causing the null standard deviation line 1).}
\label{tab:comp_natmeth}
\begin{tabular}{ccccccll}
\hline
\rule{0pt}{3ex} &\multicolumn{5}{c}{$\hat{N}-N$}& &  \\
\cline{2-6}
$v$ &  -2 & -1 & 0 & 1 & $\geq 2$ & \multicolumn{1}{c}{$\tau_1$} & \multicolumn{1}{c}{$\tau_2$} \\
\hline
0.6 & 99 & 0 & 1 & 0 & 0 & 93.0 ~ (0.0) & 177.0 ~ (0.0) \\
0.8 & 82 & 0 & 15 & 1 & 2 & 96.0 ~ (7.9) & 173.4 ~ (3.9)  \\
1.0 & 23 & 0 & 68 & 7 & 2 & 99.9 ~ (3.9) & 174.9 ~ (4.3)  \\
2.0 & 0 & 0 & 96 & 1 & 3 & 100.0 (1.4) & 175.0 ~ (1.2)  \\
\hline
\end{tabular}
\end{table}

\begin{table}[t!]
\centering
\caption[Selected models with the method of \cite{monnier2015inferring} on 100 simulated trajectories from Scenario 1]{Selected models with the method of \cite{monnier2015inferring} on 100 simulated trajectories from Scenario 1. BR (respectively BRD) stands for Brownian (respectively Brownian with drift). For instance, when $v=0.6$, the method of \cite{monnier2015inferring} states that the best fit is Brownian motion only for 97 trajectories; \BRD only for 2 trajectories; a mix of Brownian and \BRD for 1 trajectory.}
\label{app:tab:prop_mod_natmed}
\begin{tabular}{cccccc}
\hline
\rule{0pt}{3ex} $v$ & BR & BRD & BR-BR & BR-BRD & BRD-BRD \\
\hline
0.6 & 97   &  2   &  0   &  1  &   0 \\
0.8 & 74   &  8   &  0   & 18  &   0 \\
1   & 18   &  5   &  0   & 77  &   0 \\
2   & 0    &  0   &  0   & 100 &   0 \\
\hline
\end{tabular}
\end{table}

\subsection{Algorithmic Considerations}
Finally, we compare the speed and stability of the different methods. The method of \cite{monnier2015inferring} is time consuming because of the estimation of the \textit{a posteriori} distribution by the Metropolis-Hastings algorithm. Assuming $K\leq 2$, it took $115s$ in average to deal with one trajectory of the simulation presented in Table \ref{tab:comp_natmeth} (300 points) with four cores working in parallel on a Mac Book Pro version 10.10.1 equipped with 2.8 GHz Intel Core i7, 16 Gb of RAM. In comparison, our method takes less than $0.05s$ to process a trajectory without working in parallel.  Both our procedure \ref{proc_cp} and the method of \cite{turkcan2013bayesian} compute quantities on local windows (in our case the statistics \ref{eq:stat_ba}, the BIC of different models for \citep{turkcan2013bayesian}). From this aspect, the complexity of these two algorithms is equivalent. However, \cite{turkcan2013bayesian} needs to estimate the MAP (maximum \textit{a posteriori}) to compute the BIC. They choose a complex likelihood to model the spatial heterogeneity of the motion. Therefore, they use quasi-Newtonian optimization to find the MAP which is the most time consuming step of their procedure. It took in average $11 s$ to process a trajectory of the simulation presented in Table \ref{tab:comp_natmeth} (500 points) against less than $0.05s$ for Procedure \ref{proc_cp}. In term of stability, different runs of the method of \cite{monnier2015inferring} on the same trajectory can give different results (see Section \ref{sec:real_data}). This is due to a bad convergence of the Metropolis-Hastings algorithm. In rare cases, the optimization step of \cite{turkcan2013bayesian} can fail. Procedure \ref{proc_cp} does not suffer any of these problems as it does not involve any parameter inference.

\section{Real Data}
\label{sec:real_data}
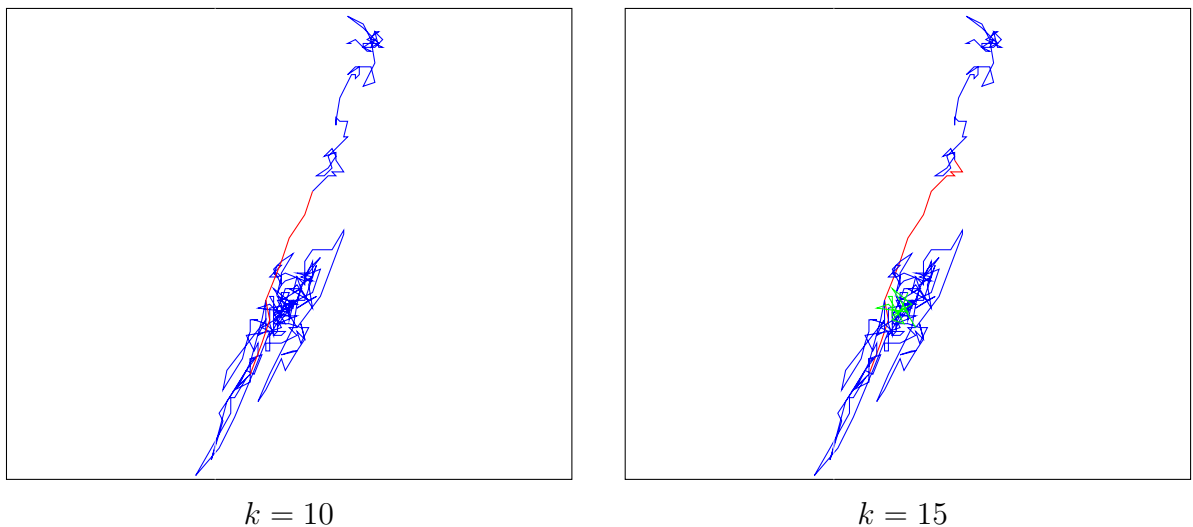
\begin{figure}[th!]
\centering
\begin{tabular}{cc}
\pgfplotstableread {fig2/traj_natmet_wind10.dat} {\loadedtable}

\begin{tikzpicture}[baseline]
\begin{axis}
[xlabel=,ylabel=,
axis equal,
minor tick num=4,
xtick=\empty,ytick=\empty,
enlarge y limits=false,
y dir = reverse,
width=0.56\linewidth,
]
\addplot[color=blue] table[x=x0,y=y0] from \loadedtable;
\addplot[color=blue] table[x=x2,y=y2] from \loadedtable; 
\addplot[color=red] table[x=x1,y=y1] from \loadedtable;

\addplot[mark=none,color=white] coordinates {
(100,105) (100,226)
};
\end{axis}
\end{tikzpicture} & \pgfplotstableread {fig2/traj_natmet_wind15.dat} {\loadedtable}

\begin{tikzpicture}[baseline]
\begin{axis}
[xlabel=,ylabel=,
axis equal,
minor tick num=4,
xtick=\empty,ytick=\empty,
enlarge y limits=false,
y dir = reverse,
width=0.56\linewidth,
]
\addplot[color=blue] table[x=x0,y=y0] from \loadedtable;
\addplot[color=red] table[x=x2,y=y2] from \loadedtable; 
\addplot[color=blue] table[x=x3,y=y3] from \loadedtable; 
\addplot[color=blue] table[x=x1,y=y1] from \loadedtable;
\addplot[color=green] table[x=x4,y=y4] from \loadedtable;
\addplot[mark=none,color=white] coordinates {
(100,105) (100,226)
};
\end{axis}
\end{tikzpicture}\\
$k=10$ & $k=15$\\
\end{tabular}
\caption[Change point detection on trajectories depicting neuronal mRNPs]{Change point detection on trajectories depicting neuronal mRNPs. The blue parts correspond to Brownian portions of the trajectory, red part to superdiffusive portions, green part to the subdiffusive portion. The detected change points are $\pmb{\tau}=(67,75)$ for $k=10$ alternating between Brownian, superdiffusion and Brownian. The detected change points are $\pmb{\tau}=(62,75,282)$ for $k=15$ alternating between Brownian motion, superdiffusion, Brownian motion and finally subdiffusion.}
\label{fig:traj_real_natmet}
\end{figure}

We use the same data as \cite{monnier2015inferring} depiciting long-range transport of mRNAs in complex with mRNA-binding proteins (mRNPs). In live neuronal cultures, endogenous $\beta$-actin mRNP particles alternate between Brownian motion and active transport. In case of active transport (superdiffusion), the particle is driven by molecular motors along microtubule tracks in the neuronal dendrites. The microscopic sequence was obtained using mRNA fluorescence labeling techniques. More specifically, in the experiment of \cite{monnier2015inferring}, the MS2 bacteriophage capsid protein was tagged with a GFP (Green Fluorescence Protein). As the MS2 bacteriophage capsid protein  binds to $\beta$-actin mRNP, it allows to track this latter.\newline
\indent The time resolution  of the sequence is $\Delta=0.1s$. The space resolution is not given but when the \BRD is chosen, \cite{monnier2015inferring} find a drift parameter with order of magnitude of $1\mu m.s^{-1}$.  
 As before, we set the parameter $K=2$ for the method of \cite{monnier2015inferring}. In this case, the model 3 (two Brownian motion with different diffusion coefficients) is selected by the method. Then, from our point of view, there are no change of dynamics.
We note that we run 100 times the algorithm and did not get the same outcome each time. It is due to the fact that the inference is based on a Monte-Carlo Markov chains (MCMC) algorithm for computing the \textit{a posteriori} estimates. Consequently, the selected model was not the same every times (92 times model 3, 7 times model 4, 1 time model 5). Then, the MCMC algorithm can show some problems of stability giving some contrary outcomes from one run to another.\newline
\indent In Figure \ref{fig:traj_real_natmet}, we show our results for two window sizes $k=10$ and $k=15$. We do not detect any change point for larger windows. With both window sizes, we detect approximately the same portion of the trajectory as superdiffusive. With the window size $k=15$, we also detect a subdiffusive part in the trajectory.


\section{Discussion}
We proposed a non parametric algorithm to detect the change points along a particle trajectory. These change points are defined as the times at which the particle switches between three modes of motion, namely Brownian motion, subdiffusion and superdiffusion. These types of processes are extensively used in the biophysic literature \cite{bressloff2014stochastic} \cite{berry2014anomalous}. When the trajectory is fully Brownian (our null hypothesis $H_0$), we control the probability to detect a false change point at level $\alpha$. Our algorithm is user-friendly as there is only one parameter to tune, namely the window size k.

 We compared our method to the method of \cite{turkcan2013bayesian} and the one of \cite{monnier2015inferring}. First, none of these methods propose to distinguish the three types of diffusion. Secondly, we show more reliable results than both of the methods on our simulations. Thirdly, our method is much faster than the two others which is a advantage when dealing with a large numbers of trajectories. We also considered real data depicting neuronal mRNPs (mRNAs in complex with mRNA-binding.
proteins).

 Future work will involve the development of a multi-scale approach to automatically select the optimal window size $k$.  
 
\bibliographystyle{agsm}

\bibliography{biblio}

\end{document}


\def\spacingset#1{\renewcommand{\baselinestretch}%
{#1}\small\normalsize} \spacingset{1}

\if1\blind
{
  \title{\bf Supplementary Materials: A Sequential Algorithm to Detect Diffusion Switching along Intracellular Particle Trajectories}
  \author{V. Briane$^{\star \ddag}$, C. Kervrann$^\star$, M. Vimond$^\ddag$\\
    INRIA, Centre de Rennes Bretagne Atlantique, Serpico Project-Team$^\star$\\
    CREST (Ensai,  Universit\'e Bretagne Loire)}
    \date{}
  \maketitle
  
} \fi

\if0\blind
{
  \bigskip
  \bigskip
  \bigskip
  \begin{center}
    {\LARGE\bf Supplementary Materials: A Sequential Algorithm to Detect Diffusion Switching along Intracellular Particle Trajectories }
\end{center}
  \medskip
} \fi

\bigskip

\label{app:proof}

\section{Proof of Proposition 1}
\begin{proof}
We suppose that the trajectory $\mathbb{X}_n$ is generated under the null hypothesis (3.2) that is the trajectory is fully Brownian. For simplicity, we note $P$ the probability under $H_0$ (noted $P_{1/2,0,\sigma}^\emptyset$ in Section (2) ).
We want to show that under $H_0$, Procedure 1 with thresholds $\gamma_1$ and $\gamma_2$ defined in Proposition 1, controls the probability of the type I error at level $\alpha$:
\begin{equation}
P\left(\exists i \in \{k,\dots,n^{\star}\}, \quad \sum\limits_{j=i}^{i+c-1}\mathbf{1}(Q_j\neq0\right)\geq c^{\star})\leq \alpha
\label{app:eq:prob_typeI}
\end{equation}
where $n^{\star}=n-k-c+1$.\newline
We express the event $\{Q_i\neq 0\}$ as:
\begin{equation}
\begin{split}
\{Q_i\neq 0\}&=\{\phi(B_i)=0,\phi(A_i)=1\}\cup\{\phi(B_i)=0,\phi(A_i)=2\}\\
&\cup\{\phi(B_i)=1,\phi(A_i)=0\}\cup\{\phi(B_i)=2,\phi(A_i)=0\}\\
&\cup\{\phi(B_i)=1,\phi(A_i)=2\}\cup\{\phi(B_i)=2,\phi(A_i)=1\}
\end{split}
\end{equation}
Then we deduce the following inclusion:
\begin{equation}
\{Q_i\neq 0\} \subset \{\phi(B_i)=1\}\cup\{\phi(B_i)=2\}\cup\{\phi(A_i)=1\}\cup\{\phi(A_i)=2\}
\label{app:eq:inclusion1}
\end{equation}
Then from the definition of $\phi$ we can reexpress the right hand side of \eqref{app:eq:inclusion1} and get the inclusion:
\begin{equation}
\begin{split}
\{Q_i\neq 0\} &\subset \{B_i<\gamma_1\}\cup\{A_i<\gamma_1\}\cup\{B_i>\gamma_2\}\cup\{A_i>\gamma_2\}\\
&=\{\min(B_i,A_i)<\gamma_1\} \cup \{\max(B_i,A_i)>\gamma_2\}
\end{split}
\end{equation}
In the sequel, we note  $d_i=\min(B_i,A_i)$ and $D_i=\max(B_i,A_i)$.
Then we have:
\begin{equation}
P(Q_i \neq 0)\leq P(\{d_i<\gamma_1\} \cup \{D_i>\gamma_2\}),\quad i=k,\dots,n^\star.
\end{equation}
This implies the following:
\begin{equation}
P\Bigl(\sum\limits_{j=i}^{i+c-1}\mathbf{1}(Q_j\neq0)\geq c^{\star}\Bigr)\leq P\Bigl(\sum\limits_{j=i}^{i+c-1}\mathbf{1}(\{d_j<\gamma_1\} \cup \{D_j>\gamma_2\})\geq c^{\star}\Bigr)
\label{app:eq:ineg_sum}
\end{equation}
Now, we can bound the right-hand side of Equation \eqref{app:eq:ineg_sum}:
\begin{equation}
\begin{split}
&P\Bigl(\sum\limits_{j=i}^{i+c-1}\mathbf{1}(\{d_j<\gamma_1\} \cup \{D_j>\gamma_2\})\geq c^{\star}\Bigr)\\
\leq & P\Bigl(\sum\limits_{j=i}^{i+c-1}\mathbf{1}(\{d_j<\gamma_1\}) + \mathbf{1}(\{D_j>\gamma_2\})\geq c^{\star}\Bigr)\\
\leq & P\Bigl(\sum\limits_{j=i}^{i+c-1}\mathbf{1}(\{d_j<\gamma_1\})\geq c^{\star}/2\Bigr) + P\Bigl(\sum\limits_{j=i}^{i+c-1}\mathbf{1}(\{D_j>\gamma_2\})\geq c^{\star}/2\Bigr)
\end{split}
\label{app:eq:ineg_sum2}
\end{equation}
Then, we can express the right-hand side of Equation \eqref{app:eq:ineg_sum2} as:
\begin{equation}
\begin{split}
&P\Bigl(\sum\limits_{j=i}^{i+c-1}\mathbf{1}(\{d_j<\gamma_1\})\geq c^{\star}/2\Bigr) +P\Bigl(\sum\limits_{j=i}^{i+c-1}\mathbf{1}(\{D_j>\gamma_2\})\geq c^{\star}/2\Bigr)\\
=&P(d_{i(c^\star/2)}<\gamma_1)+P(D_{i(c-c^\star/2)}>\gamma_2)
\label{app:eq:ineg_sum3}
\end{split}
\end{equation}
Finally we have:
\begin{equation}
\begin{split}
&P(\exists i \in \{k,\dots,n^\star\}, \quad \sum\limits_{j=i}^{i+c-1}\mathbf{1}(Q_j\neq0)\geq c^{\star})\\
=&P\Bigl(\bigcup_{i=k}^{n^\star}\Bigl\{\sum\limits_{j=i}^{i+c-1}\mathbf{1}(Q_j\neq0)\geq c^{\star}\Bigr\}\Bigr)\\
\leq & P\Bigl( \bigcup_{i=k}^{n^\star} \Bigl\{\sum\limits_{j=i}^{i+c-1}\mathbf{1}(\{d_j<\gamma_1\})\geq c^{\star}/2\Bigr\}\Bigr) + P\Bigl(\bigcup_{i=k}^{n^\star} \Bigl\{\sum\limits_{j=i}^{i+c-1}\mathbf{1}(\{D_j>\gamma_2\})\geq c^{\star}/2\Bigr\}\Bigr)\\
=& P\Bigl(\bigcup_{i=k}^{n^\star} \{d_{i(c^\star/2)}<\gamma_1\}\Bigr)+P\Bigl( \bigcup_{i=k}^{n^\star} \{D_{i(c-c^\star/2)}>\gamma_2\}\Bigr)\\
=& P\Bigl(\min_{i=k,\dots,n^\star} d_{i(c^\star/2)}<\gamma_1\Bigr)+P\Bigl( \max_{i=k,\dots,n^\star} D_{i(c-c^\star/2)}>\gamma_2\Bigr)\\
=&\frac{\alpha}{2}+\frac{\alpha}{2}=\alpha
\end{split}
\end{equation}
We go from line 2 to line 3 using Equations \eqref{app:eq:ineg_sum} and \eqref{app:eq:ineg_sum2}. We go from line 3 to line 47 using Equation \eqref{app:eq:ineg_sum3}. Finally, we go from line 5 to 6 using the thresholds $\gamma_1$ and $\gamma_2$ of Proposition 1. It finishes the proof.
\end{proof} 
\newpage
\section{Monte Carlo Algorithm for Estimating $(\gamma^\star_1,\gamma^\star_2)$}

\begin{algorithm}[ht!]
\KwIn{$n$, $k$, $\alpha,$ $N$}
\tcp{the length $n$ of the trajectory}
\tcp{the size $k$ of the subtrajectories}
\tcp{the level $\alpha\in(0,1)$}
\tcp{the number $N$ of Monte Carlo experiments }
\KwResult{$\hat{\gamma}_1(n,k,\alpha)$ $\hat{\gamma}_2(n,k,\alpha)$}
\For{i=1 \emph{\KwTo} N}{
Generate $\mathbb{X}_n^i$ of size $n$ from the null hypothesis (3.2) (see the paper)  with $\sigma=1$ and $\Delta=1$ \;
\tcp{Compute the statistics (4.1) (see the paper) along $\mathbb{X}_n^i$} 
\For{j=k \emph{\KwTo} n-k}{
Compute $(B^i_j,A^i_j)$ from (4.1)\;
Set $d^i_j=\min(B^i_j,A^i_j)$\;
Set $D^i_j=\max(B^i_j,A^i_j)$\;
}
\For{r=k \emph{\KwTo} n-k-c+1}{
Compute $s^i_r$ the $c^{\star}/2$ smallest element of $(d^i_r,\dots,d^i_{r+c-1})$\;
Compute $S^i_r$ the $c-c^{\star}/2$ smallest element of $(d^i_r,\dots,d^i_{r+c-1})$\;
}
Compute $m_i=\min_r(S^i_r)$ and $M_i=\max_r(s^i_r)$\;
}
Let $(\tilde{m}_1,\dots,\tilde{m}_N)$ the sorted $m_i$s and $(\tilde{M}_1,\dots,\tilde{M}_N)$ the sorted $M_i$s\;
Set $\hat{\gamma}_1(n,k,\alpha)=\tilde{m}_{\lfloor(\alpha/2)N\rfloor}$ and $\hat{\gamma}_2(n,k,\alpha)=\tilde{M}_{\lfloor(1-\alpha/2)N\rfloor}$\;
\vspace{0.5cm}
\caption[]{Estimation of the cut-off values $\gamma^\star_1$ and $\gamma^\star_2$ by Monte Carlo simulations. For estimating ($\tilde{\gamma}_1,\tilde{\gamma}_2$), one should turn $c^\star/2$ into $c^\star$ in this algorithm. }
\label{algo:cut_off_val}
\end{algorithm} 

\bibliographystyle{agsm}
\bibliography{biblio.bib}